\documentclass[runningheads]{llncs}

\usepackage{graphicx}
\usepackage{hyperref}
\usepackage{amsfonts}
\begin{document}

\title{TEDS-Net: Enforcing Diffeomorphisms in Spatial Transformers to Guarantee Topology Preservation in Segmentations}
\titlerunning{Enforcing Diffeomorphisms to Preserve Topology in Segmentations}
%
\author{Madeleine K. Wyburd\inst{1}, Nicola K. Dinsdale \inst{2}, Ana I.L. Namburete$^{\dagger}$ \inst{1} \and Mark Jenkinson$^{\dagger}$ \inst{2,3,4}}

\authorrunning{M.K. Wyburd et al.}

\institute{Institute of Biomedical Engineering, Department of Engineering Science, University of Oxford, Oxford, UK\\
\and
Wellcome Centre for Integrative Neuroimaging, FMRIB,University of Oxford, Oxford, UK\\
\and
Australian Institute for Machine Learning (AIML), Department of Computer Science, University of Adelaide, Adelaide, Australia\\
\and
South Australian Health and Medical Research Institute (SAHMRI), North Terrace, Adelaide, Australia\\
\email{madeleine.wyburd@hertford.ox.ac.uk}\\
}

\maketitle  
\begin{abstract}
Accurate topology is key when performing meaningful anatomical segmentations, however, it is often overlooked in traditional deep learning methods. In this work we propose TEDS-Net: a novel segmentation method that guarantees accurate topology. Our method is built upon a continuous diffeomorphic framework, which enforces topology preservation. However, in practice, diffeomorphic fields are represented using a finite number of parameters and sampled using methods such as linear interpolation, violating the theoretical guarantees. We therefore introduce additional modifications to more strictly enforce it. Our network learns how to warp a binary prior, with the desired topological characteristics, to complete the segmentation task. We tested our method on myocardium segmentation from an open-source 2D heart dataset. TEDS-Net preserved topology in 100$\%$ of the cases, compared to 90$\%$ from the U-Net, without sacrificing on Hausdorff Distance or Dice performance. Code will be made available at: \url{www.github.com/mwyburd/TEDS-Net}.

\end{abstract}

\section{Introduction}


Anatomical segmentations with incorrect topology can be clinically problematic and highly impractical. For example, when segmenting the heart's myocardium, topological mistakes may present as isolated segmented regions, small holes within the wall or a disconnected perimeter. These inaccuracies can all lead to incorrect measurements of circumference and wall thickness, which are often required when diagnosing heart conditions such as hypertrophic cardiomyopathy \cite{marian2017hypertrophic}. Post-processing morphological operations, such as binary closing, can be used to correct some small gaps and disconnected regions. However, they are unaware of the structure's overall topology and need to be customised accordingly, therefore, an unified method is preferred.

Semantic segmentation of medical images is regularly performed using deep convolutional neural networks (CNNs), such as the U-Net \cite{ronneberger2015u}. One alternative segmentation method uses a combination of CNNs and spatial transformers to learn the spatial warping required to transform a set of prior shapes into the desired class labels \cite{jaderberg2015spatial,wickramasinghe2019probabilistic,vigneault2018omega,dinsdale2019spatial,zeng2019liver}. Such methods have outperformed conventional encoder-decoder and state-of-the-art architectures, however, they have no topological guarantees. 

Moreover, both these methods are commonly trained and evaluated using loss functions such as Dice or binary cross-entropy, which evaluate each pixel individually and not the higher level structure, often resulting in large topological errors. Recent work has recognised the importance of accurate whole-structure segmentation, with the development of loss functions that encourage the preservation of topology \cite{clough2020topological,mosinska2018beyond,hu2019topology}. Although these techniques report an improved topology accuracy compared to standalone pixel-wise loss functions, they only encourage topology preservation as opposed to enforcing it, with the best performing method achieving 96.7\% topology accuracy in myocardium segmentation \cite{clough2020topological}.

In theory, topology will always be preserved by deforming a prior, with the correct topological characteristic, with a continuous \textit{diffeomorphic field} \cite{ashburner2007fast}. Diffeomorphic fields are continuous deformation fields that result in a one-to-one mapping. Their derivatives are invertible, resulting in positive Jacobian determinants, giving an unambiguous mapping between coordinates and therefore preserving topology \cite{ashburner2007fast}. 

Resampling one diffeomorphic field by another, also known as a \textit{composition}, always results in a new diffeomorphic field \cite{ashburner2007fast}. In VoxelMorph's methods \cite{dalca2018unsupervised,dalca2019unsupervised}, this elegant property has been utilised for brain registration, by initialising with a small diffeomorphic field and amplifying it through a series of novel integration layers, based on the squaring and scaling approach \cite{arsigny2006log,moler2003nineteen}, to generate a topology-preserving large-deformation field. However, when diffeomorphisms are applied in discrete settings, such as images stored as discrete voxels, their topological guarantees can begin to break down. This is often indicated by the emergence of non-positive Jacobian determinants in the deformation fields, which correspond to folding voxels in the warped space \cite{dalca2019unsupervised}. In brain registration a small fraction of violations may be manageable, conversely, for segmentation tasks these can lead to topological errors in the structure of interest, defeating the purpose. Here, we adapt VoxelMorph's methods to be suitable for segmentation tasks by introducing a novel topology-preserving activation function and additional smoothing terms to counteract the negative effects of discrete sampling. These modifications enforce true diffeomorphisms, shown to have only positive Jacobian determinants, which results in 100\% topology preservation.

In this work, we present the novel Topology Enforcing Diffeomorphic Segmentation Network (TEDS-Net), which to the best of our knowledge is the first deep learning technique to achieve 100\% topology accuracy, and to combine spatial transformer networks (STN) and diffeomorphic displacement fields to complete a segmentation as the primary task. Our method uses CNNs to learn a warp field that maps a binary prior on to the input image, with the warp field being diffeomorphic by construction. We show that in the discrete setting, continuous properties of diffeomorphisms are no longer guaranteed, therefore, we introduce model components that enforce topology preservation in all cases.

%
\section{Method}
\begin{figure}[t]
    \centering
    \includegraphics[width=0.9\textwidth]{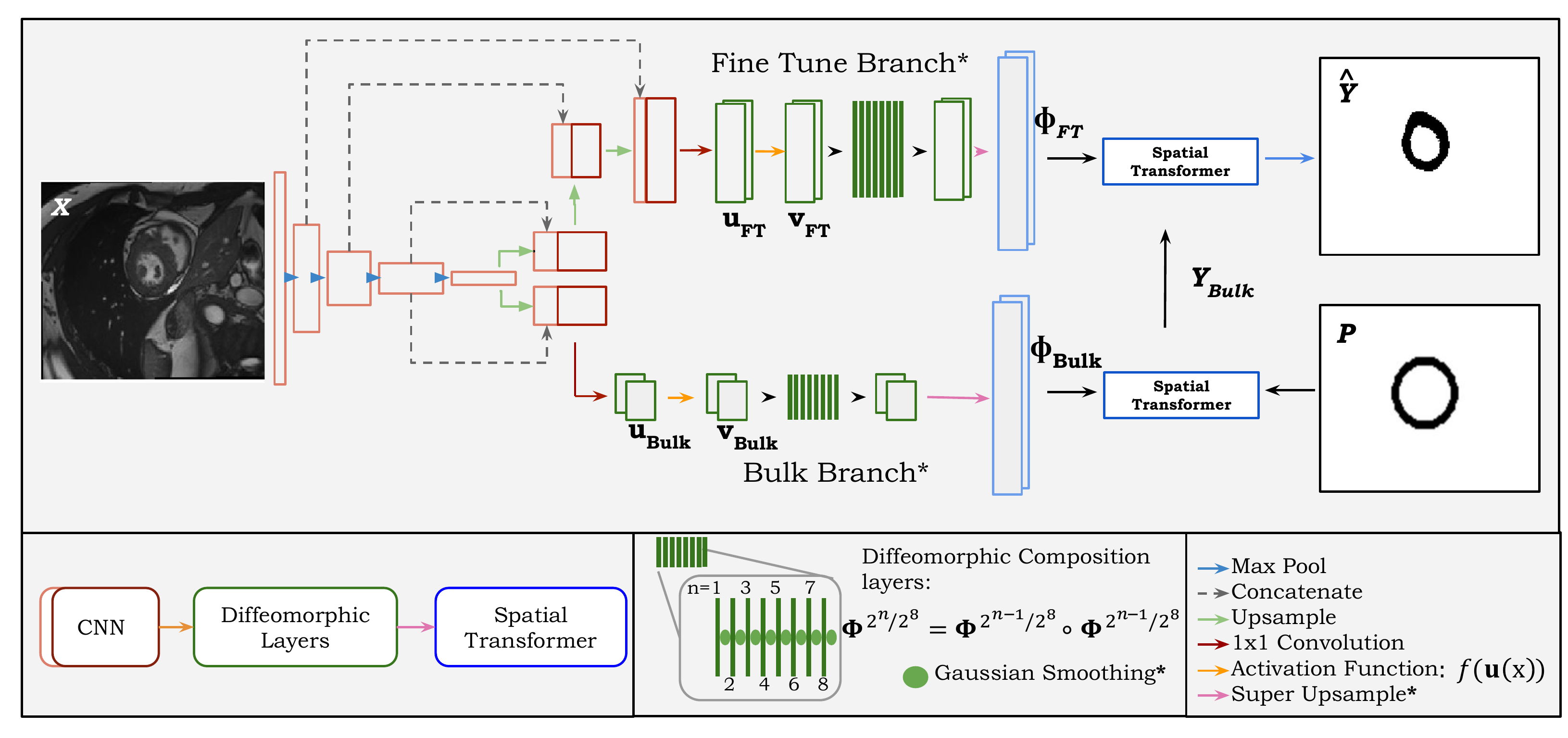}
    \caption{Schematic of TEDS-Net architecture. A CNN learns two initial fields, $\textbf{u}$, at different resolutions, from an input image $\textbf{X}$. The fields are enforced to be diffeomorphic using an activation function, $\textbf{v}= f(\textbf{u})$, amplified through composition layers  and ``super" upsampled to 2x the resolution of the input. The bulk displacement, $\mathbf{\Phi}_{\mathrm{Bulk}}$, samples a binary prior, $\mathbf{P}$, generating $\mathbf{Y}_{\mathrm{Bulk}}$, which is then sampled by the fine tuning field, $\mathbf{\Phi}_{\mathrm{FT}}$. The asterisks show the elements removed during ablation studies.}
    \label{fig:Arch}
\end{figure}

The aim of the proposed TEDS-Net is to predict a segmentation label, $\hat{\mathbf{Y}}$, using learnt diffeomorphic transformations, $\mathbf{\Phi}$, applied to a binary prior image, $\mathbf{P}$. The network is comprised of three main parts, shown in the bottom left corner of Fig \ref{fig:Arch}. Firstly, an encoder-decoder style network is used to learn the initial velocity fields. These fields are then enforced to be diffeomorphic and amplified using the squaring and scaling approach \cite{moler2003nineteen,arsigny2006log}. Finally, the diffeomorphic fields are used to sample a prior binary shape to complete the segmentation task.

\subsubsection{CNN.}
An encoder-decoder network was used to extract the relevant features from an input image, $\mathbf{X}\in\mathbb{R}^{h\times w}$ with dimensions $[h \times w]$, in order to predict two initial velocity fields: a \textit{bulk field} and a \textit{fine-tuning field}. We used an architecture similar to the U-Net but with two decoder branches \cite{ronneberger2015u,stergios2018linear}. Each branch consisted of a series of convolutions followed by an instance normalisation and ReLU activation function and then repeated, referred to as a convolutional block. 
 



Branching off at the bottleneck were two decoder streams, made up of convolutional blocks and a final 1x1 convolution, used to generate a two-dimensional field, $\textbf{u}$. The first stream, with one upsampling convolutional block, predicts a low resolution bulk-velocity, $\textbf{u}_{\mathrm{Bulk}}\in\mathbb{R}^{2 \times (h \times w)/8}$, which focuses on warping the prior shape into the correct region. The second, with three upsampling blocks, predicts a higher resolution fine-tuning field, used to fine-tune the warped shape,  $\textbf{u}_{\mathrm{FT}}\in\mathbb{R}^{2 \times (h \times w)/2}$. 


\subsubsection{Diffeomorphic Layers.}
To enforce that the initial fields, $\mathbf{u}$, were diffeomorphic, and therefore suitable for the scaling and squaring approach, a customised tanh activation function, $f(\mathbf{u})$, was applied to the output of each decoder branch to enforce that the displacements were between $-0.5$ voxel and $+0.5$ voxel and therefore topology preserving:
 

\begin{equation}\label{eq:act}
    \mathbf{v}(\mathbf{x})=f(\mathbf{u}(\mathbf{x})) = 0.5\left ( \frac{e^{\mathbf{u}(\mathbf{x})}-e^{\mathbf{-u}(\mathbf{x})}}{e^{\mathbf{u}(\mathbf{x})}+e^{\mathbf{-u}(\mathbf{x})}}\right ).
\end{equation}

A series of composition layers were then used to amplify the initial diffeomorphic fields, $\mathbf{v}$, adapted from the VoxelMorph's implementation \cite{dalca2018unsupervised,dalca2019unsupervised}. 

\paragraph{Gaussian Smoothing:}
As the diffeomorphic fields are represented using a finite number of parameters and sampled using linear interpolation, the theoretical guarantees are violated, risking key properties such as topological preservation. Moreover, these inaccuracies and imperfections can be amplified with the number of compositions performed \cite{ashburner2007fast}. To reduce these violations, we introduced Gaussian smoothing between each integration layer. For this work we used a 5 by 5 kernel and $\sigma=2$ and the impact of this will be investigated in future work. 


\paragraph{Super Upsample:}
To further smooth the fields and minimise any abnormalities, the amplified diffeomorphic flow fields were both ``super upsampled" to double the resolution of the input image, $\mathbf{X}$, using bilinear interpolation.

\subsubsection{Spatial Transformers.}
\begin{figure}[t]
    \centering
    \includegraphics[width=\textwidth]{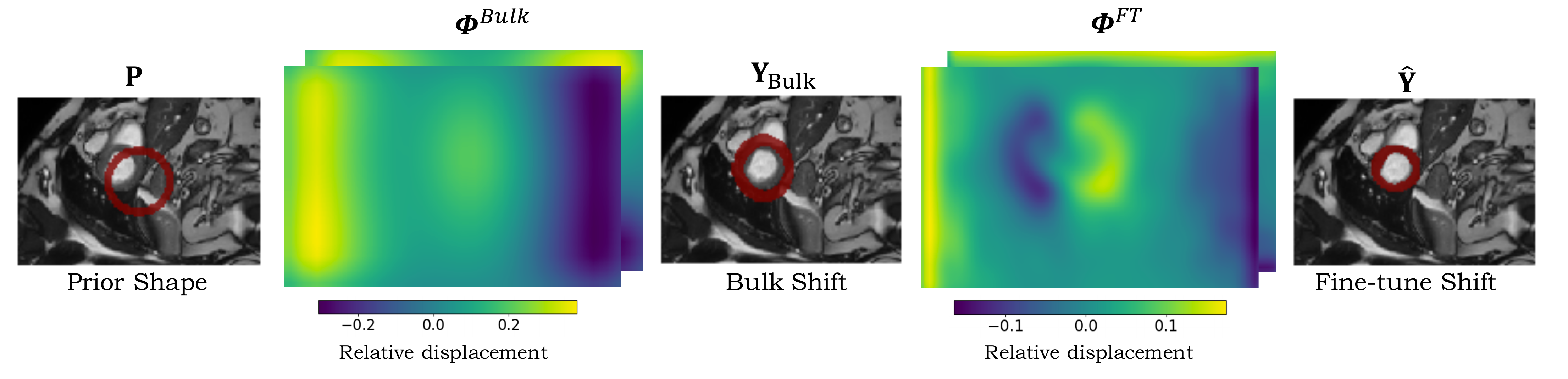}
    \caption{A prior is sampled with a displacement field at double the resolution, $\mathbf{Y}_{\mathrm{Bulk}} =  \mathbf{\Phi}_{\mathrm{Bulk}}(\mathbf{P})$, before the second super upsampled displacement is applied to the shifted image $\hat{\mathbf{Y}}=\mathbf{\Phi}_{\mathrm{FT}}(\mathbf{Y}_{\mathrm{Bulk}})$. The  displacement panels show the shift in both directions.}
    
    \label{fig:BulkFT}
\end{figure}

The diffeomorphic field, $\mathbf{\Phi}_{\mathrm{Bulk}}\in\mathbb{R}^{2 \times 2h \times 2w}$, generated from the first decoder branch sampled the prior shape, $\mathbf{P}\in\mathbb{R}^{h \times w}$, using bilinear interpolation: $\mathbf{Y}_{\mathrm{Bulk}} =  \mathbf{\Phi}_{\mathrm{Bulk}}(\mathbf{P})$, as shown in Fig \ref{fig:BulkFT}. The resulting warped image was then sampled by the fine-tuning field generated by the second decoder branch, $\mathbf{\hat{Y}}=\mathbf{\Phi}_{\mathrm{FT}}(\mathbf{Y}_{\mathrm{Bulk}})$, where $\mathbf{\Phi}_{\mathrm{FT}}\in\mathbb{R}^{2 \times 2h \times 2w}$. The prediction, $\mathbf{\hat{Y}}$, was then downsampled to the same resolution as the label, $\textbf{Y}\in\mathbb{R}^{h \times w}$, using Max Pooling.

\paragraph{Field Regularisation:}
The network outputs the final label prediction, $\hat{\mathbf{Y}}$, and the two diffeomorphic transformation fields: $\mathbf{\Phi}_{\mathrm{Bulk}},\mathbf{\Phi}_{\mathrm{FT}}$, which were all used for the training loss. A combination of Dice Loss ($\mathcal{L}_{\mathrm{Dice}}$) and a field regularisation ($\mathcal{L}_{\mathrm{Grad}}$) function were used for training:

\begin{equation}\label{eq:loss}
    \mathcal{L} = \mathcal{L}_{\mathrm{Dice}}(\mathbf{Y},\hat{\mathbf{Y}})  + \beta \mathcal{L}_{\mathrm{Grad}}(\mathbf{\Phi}_{\mathrm{FT}})+ \beta \mathcal{L}_{\mathrm{Grad}}(\mathbf{\Phi}_{\mathrm{Bulk}}),
    \label{eq:loss}
\end{equation}

where $\mathcal{L}_{\mathrm{Grad}} = \sum_{i,j=1}^{2h,2w}\left \| \nabla  \mathbf{\Phi}(i,j) \right \|^{2} $, which encourages smooth flow fields by penalising large spatial gradients between neighbouring voxels, adapted from \cite{balakrishnan2019voxelmorph}. A weighting parameter, $\beta$, was used to balance the contributions between $\mathcal{L}_{\mathrm{Grad}}$ and $\mathcal{L}_{\mathrm{Dice}}$, in this work we used a constant $\beta$ of $10,000$, in order to give all the loss function components similar magnitudes, as measured empirically.

\subsection{Experimental Setup}

TEDS-Net was used to segment the myocardium from an open source dataset of MRI heart slices in the short axis\footnote{The ACDC database: \url{www.creatis.insa-lyon.fr/Challenge/acdc}} \cite{bernard2018deep}. This dataset consisted of 2 scans from 100 patients, across five different pathologies. Five myocardium-containing slices were taken from each scan, cropped to 144 by 208 pixels and augmented using rotations, shifts, zooming and a combination of the three, resulting in 8,000 images split 75\%, 15\% and 10\% for training, validation and testing, respectively. It was assured that slices from the same patient were assigned to the same subset.

The topology of the myocardium is equivalent to a hollow circle, which we used as the prior shape, $\mathbf{P}$, shown in Fig \ref{fig:BulkFT}. It should be noted that the patients' scans were not aligned and therefore the location of the myocardium was inconsistent. Additionally, different pathologies have varying myocardium thickness. Despite this, the same arbitrary prior shape was used throughout, illustrating the robustness of the method.

Two main experiments were completed and evaluated on 100 unseen heart slices without augmentations. Firstly, ablation studies were performed to show the contribution of each additional element used in TEDS-Net to tackle the limitations of diffeomorphic sampling, shown with the asterisks in Fig \ref{fig:Arch}, using a prior shape of radius 30 and 8 integration layers, which were empirically set. As a comparison, we used a U-Net \cite{ronneberger2015u} and VoxelMorph diffeomorphic network \cite{dalca2018unsupervised,dalca2019unsupervised}, using the shape prior $\textbf{P}$ as the second input channel. Secondly, we investigated the effect of the number of integration layers and radius of the binary prior.

The networks were all trained end-to-end on a NVIDIA GeForce GTX 1080 GPU and implemented in Pytorch using Python 3.6. Training was performed over 200 epochs, using the Adam optimiser with a learning rate of 0.0001 and a batch size of 5. TEDS-Net and the U-Net were both made up of 5 layers, with 12 initial feature maps, and a 20\% dropout applied to each map. 

\section{Results and Discussion}

To evaluate segmentation performance, Dice, Hausdorff Distance (HD), and a count of incorrect topologies were used, as seen in Table \ref{tab:Results}. A predicted segmentation's topology was classed as incorrect if their Betti numbers \cite{millson1976first} varied from the known topological properties of the myocardium. 
To assess the discrete fields, the Jacobian determinants were also measured, as perfect diffeomorphic fields always have strictly positive determinants.



\begin{table}[t]
\renewcommand{\arraystretch}{1.5}
\centering
\caption{Comparison of myocardium segmentation performed by the baselines and TEDS-Net, accompanied by ablation studies (A1-6). The mean and standard deviation are given for Dice, Hausdorff Distance (HD) and percentage of non-positive Jacobian determinants from each field. The best result from each measure is shown in bold.}

\label{tab:Results}
\resizebox{\textwidth}{!}{%
\begin{tabular}{lll|l|l|l}
\hline
\multicolumn{1}{l|}{\textbf{Network}}                                                  & \multicolumn{1}{l|}{\textbf{Dice}}            & \textbf{HD}              & \textbf{$\%\left | J_{\mathbf{\Phi}_{\mathrm{Bulk}}} \right|\leq0 $} & \textbf{$\%\left | J_{\mathbf{\Phi}_{\mathrm{FT}}} \right| \leq 0$} & \multicolumn{1}{l}{\textbf{\begin{tabular}[c]{@{}l@{}}Incorrect \\ Topology\end{tabular}}} \\ \hline \hline
\multicolumn{1}{l|}{\textbf{U-Net} (baseline)}                                    & \multicolumn{1}{l|}{\textbf{$\mathbf{0.87 \pm 0.16}$}} & $5.25 \pm 13.11$         & N/A                                      & N/A                                    & \multicolumn{1}{l}{10/100}                      \\ \hline
\multicolumn{1}{l|}{\textbf{VoxelMorph} (baseline)}                                    & \multicolumn{1}{l|}{$0.79 \pm 0.18$} & $3.0 \pm 2.4$         & N/A                                      & $6.7e^{-05}\pm 0.7e^{-03} $                                    & \multicolumn{1}{l}{6/100}                      \\ \hline\hline


\multicolumn{1}{l|}{\textbf{(A1) TEDS-Net:} No Gaussian, $\mathcal{L}_{\mathrm{Grad}}$ or Super Upsample}                         & \multicolumn{1}{l|}{$0.82 \pm 0.19 $}          & $4.26 \pm 5.76$          & $56.32 \pm 1.97$                              & $43.8 \pm 3.46$                         & \multicolumn{1}{l}{17/100}              \\ \hline

\multicolumn{1}{l|}{\textbf{(A2) TEDS-Net:} No Gaussian Smoothing}                       & \multicolumn{1}{l|}{$0.86\pm 0.13$}          & $3.73 \pm 8.34$          & $67.10 \pm 7.07$                         & $35.81 \pm 4.41$                       & \multicolumn{1}{l}{3/100}                       \\ \hline

\multicolumn{1}{l|}{\textbf{(A3) TEDS-Net:} No $\mathcal{L}_{\mathrm{Grad}}$} & \multicolumn{1}{l|}{$0.85 \pm 0.14$}          & $2.77 \pm 2.01$          & $5.94 \pm 2.57$                              & $22.15 \pm 3.67$                        & \multicolumn{1}{l}{35/100}                       \\ \hline

\multicolumn{1}{l|}{\textbf{(A4) TEDS-Net:} No Super Upsample}                         & \multicolumn{1}{l|}{$0.85\pm0.11$}          & $2.89 \pm 1.61$          & $\mathbf{0.0 \pm 0.0}$                              & $\mathbf{0.0 \pm 0.0}$                         & \multicolumn{1}{l}{\textbf{0/100}}              \\ \hline 

\multicolumn{1}{l|}{\textbf{(A5) TEDS-Net:} Only Bulk Branch}                 & \multicolumn{1}{l|}{$0.77 \pm 0.15$}           & $3.86 \pm 3.50$           & $0.03 \pm 0.17$                             & N/A                                    & \multicolumn{1}{l}{\textbf{0/100}}              \\ \hline
\multicolumn{1}{l|}{\textbf{(A6) TEDS-Net:} Only Fine-Tune Branch}             & \multicolumn{1}{l|}{$0.85 \pm 0.14$}           & $3.01 \pm 2.16$           & N/A                                      & $\mathbf{0.0 \pm 0.0}$                          & \multicolumn{1}{l}{\textbf{0/100}}              \\ \hline \hline

\multicolumn{1}{l|}{\textbf{TEDS-Net} (ours)}                                   & \multicolumn{1}{l|}{$0.86 \pm 0.12$}          & {$\mathbf{2.76 \pm 2.13}$} & $\mathbf{0.0 \pm 0.0}$                            & $\mathbf{0.0 \pm 0.0}$                          & \multicolumn{1}{l}{\textbf{0/100}}              \\ \hline
\end{tabular}%
}

\end{table}

\begin{figure}[h]
    \centering
    \includegraphics[width=0.9\textwidth]{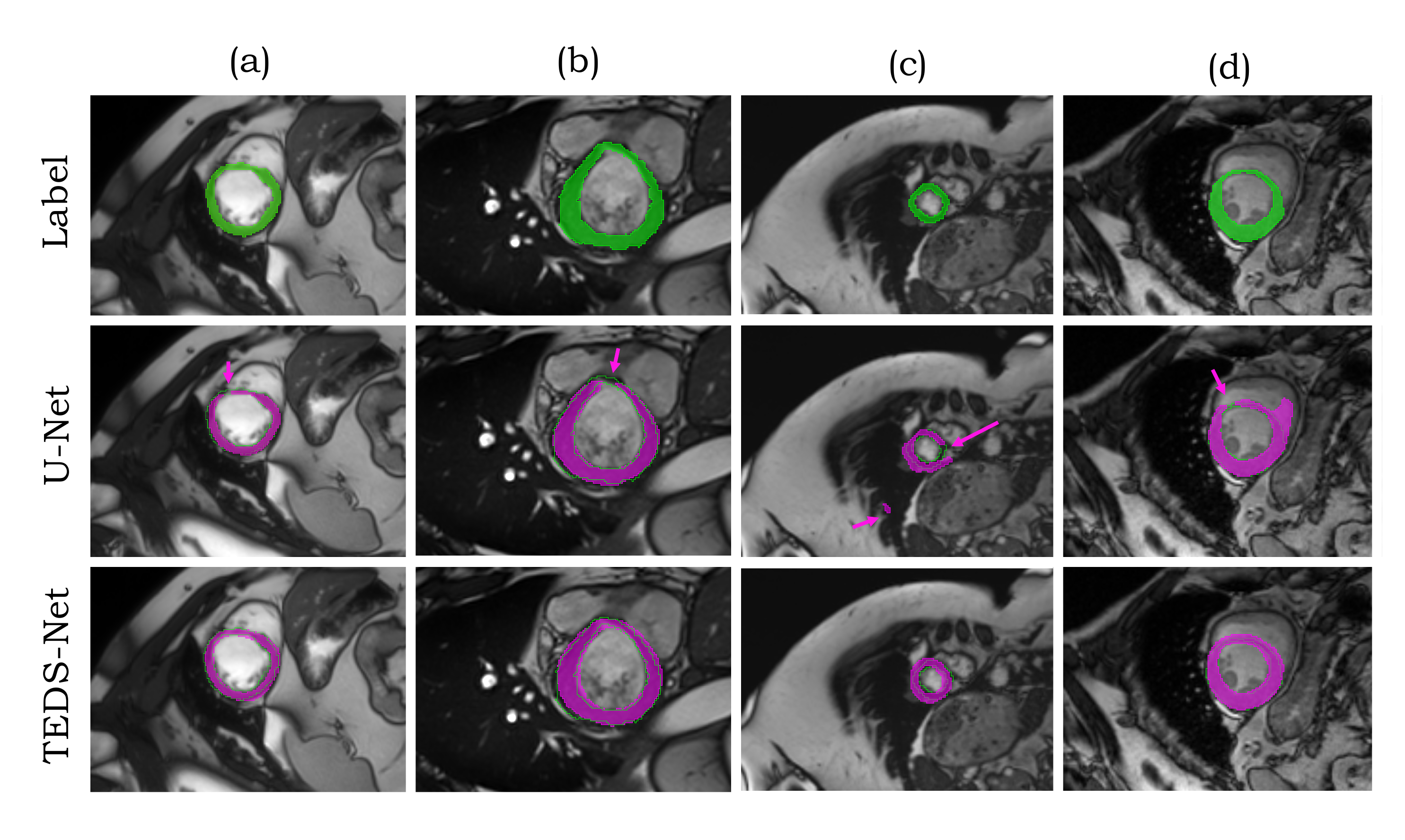}
    \caption{Examples of myocardium manual annotations (green), compared to U-Net and TEDS-Net segmentations (pink) from the ACDC dataset \cite{bernard2018deep}.}
    \label{fig:myo}
\end{figure}

Large topological errors were found when segmenting the myocardium with the U-Net, as seen from Table \ref{tab:Results}. These errors were mainly expressed as large gaps separating the myocardium, shown in Fig \ref{fig:myo}, making clinical measures of circumference extremely challenging. VoxelMorph was designed to maximise registration performance using diffeomorphisms, however, when applied to segmentation a small fraction of topology errors are observed, shown in Table \ref{tab:Results}: VoxelMorph. Unlike the U-Net predictions, these topology violations are seen at the boundaries of the myocardium, forming holes or disconnected regions when defined using 4-pixel connectivity, as done here, but not when using 8-pixel connectivity. Conversely, our TEDS-Net enforced diffeomorphisms, preserving topology for all cases, which to the best of our knowledge, is the first deep learning segmentation technique to achieve 100\% topological accuracy. Therefore, TEDS-Net additional modifications are required to prioritise topology preservation.



Paired t-tests were computed between TEDS-Net and U-Net and found that TEDS-Net significantly outperformed the U-Net in Hausdorff Distance, with a p-value of 0.01. Although the Dice accuracies were competitive, the U-Net performed significantly better (p=0.04). 
Figure \ref{fig:myo}c-d shows particularly challenging examples, where the U-Net fails to return the correct topology. However, although TEDS-Net returns labels with accurate topology, the myocardial wall appears too thick in some parts. This is likely due to the smoothing modification, whose parameter will be further investigated in future work.

Ablation studies were performed to show the effect of the additional smoothing modifications used to encourage perfect diffeomorphisms, as shown in Table \ref{tab:Results} (A1-6). The continuous guarantees of diffeomorphic sampling, such as topology preservation, break down when applied in the discrete setting, shown by the emergence of non-positive Jacobian determinants that can then lead to topological errors. There are two sources of such violations in TEDS-Net:  the composition layers and the image sampling.

To limit the numerical inaccuracies brought about by the discrete composition of two diffeomorphic fields, we introduced Gaussian smoothing between each layer. Without this addition, a large fraction of the Jacobian determinants no longer remained positive, corresponding to a number of topological defects, as shown in Table \ref{tab:Results} (A2). 

Sampling a discrete binary image with a finite warp field often results in disconnects, due to interpolations and the use of thresholds on the resultant image. To reduce this effect, we regularised the smoothness of the final deformation fields with $\mathcal{L}_{\mathrm{Grad}}$. Without this term, 35 out of 100 images were found to have topological defects and therefore, it played a vital role in mimicking perfect diffeomorphisms, as shown in Table \ref{tab:Results} (A3). 

Removing either the bulk or fine-tuning sampling branch was found to reduce the performance of both Dice and HD, as shown Table \ref{tab:Results} (A5-6). As the images are unaligned, with varying sizes of myocardium, shown in Fig \ref{fig:myo}, both branches are required to first align the prior before fine-tuning the warped shape. TEDS-Net performs the worst without the fine-tuning branch, which is likely due to the lack of flexibility in the bulk transformation, as the deformation fields are generated at a much lower resolution before being upsampled. 


\begin{figure}[t]
    \centering
    \includegraphics[width=0.9\textwidth]{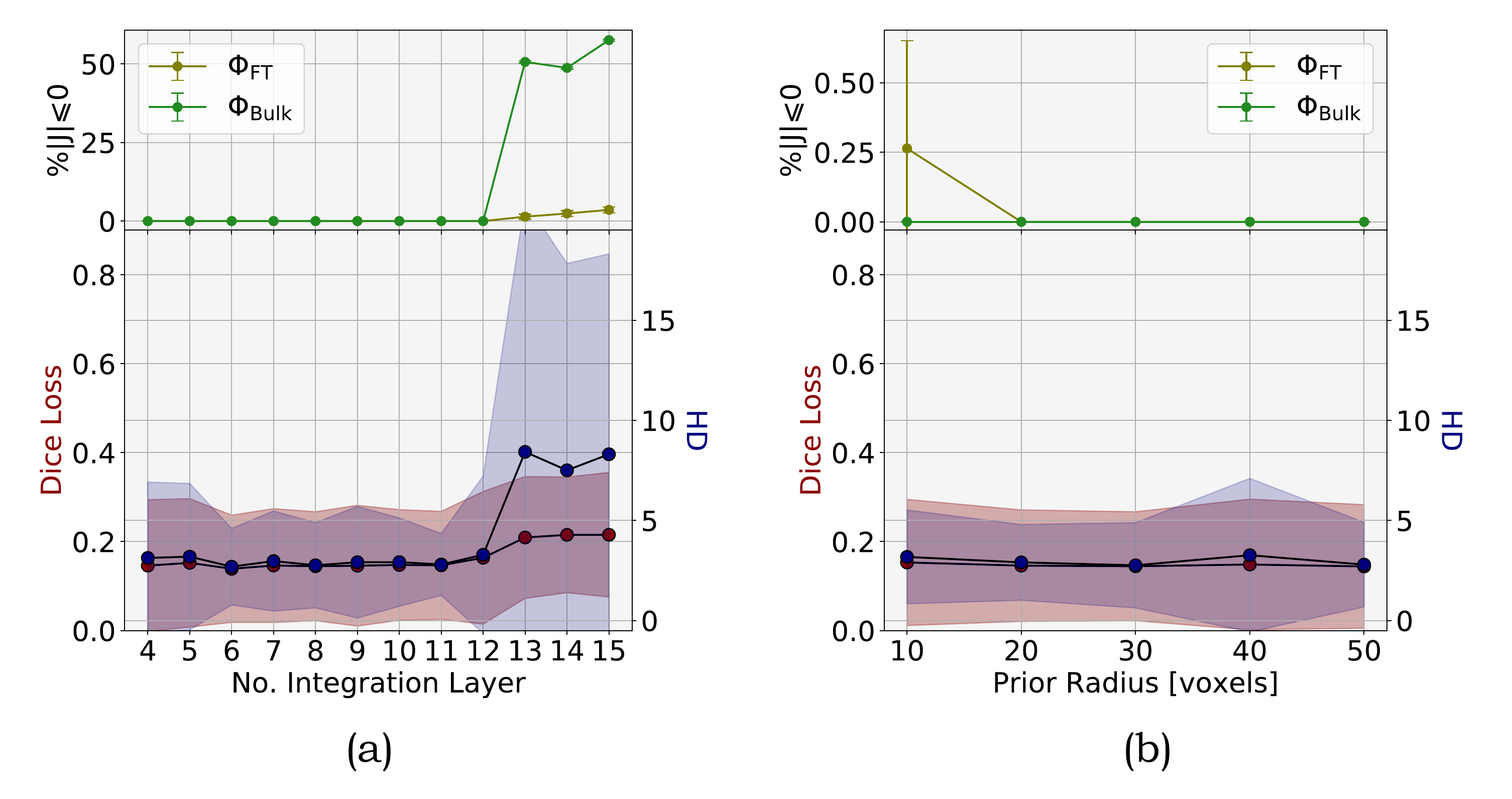}
    \caption{The effect that the number of integration layers (a) and the radius of the binary prior (b) had on segmentation performance and the diffeomorphic nature of the generated fields. Due to the image dimensions, 50 was the maximum radius used.}
    \label{fig:intlayers_PS}
\end{figure}

\subsubsection{Number of Integration Layers:}
Including Gaussian smoothing between the integration layers played a key role in enforcing topology preservation, as shown in Table \ref{tab:Results} (A2). Although theoretically accuracy should increase with the number of composition layers, this requires performing more compositions that can each bring about small violations, due to the discrete nature of the fields \cite{ashburner2007fast}. However, limiting the number of composition layers, limits the size of deformations. To investigate this further, we varied the number of integration layers used in TEDS-Net and measured the segmentation performance and Jacobian determinants of the resulting fields, shown by Fig \ref{fig:intlayers_PS}a. When using 12 or fewer integration layers, the segmentation performance is stable whilst the Jacobian determinants all remain positive and all topology is preserved. However, when the number of layers is increased beyond 12, non-positive Jacobian determinants emerge. Therefore, whilst Gaussian smoothing has been shown to be essential in the integration layers, there is a limit to the number of compositions that can be used whilst enforcing diffeomorphic sampling. 

\subsubsection{Prior Radius:}
In the test set, the myocardium's radius ranged between approximately 10 to 60 voxels, so we investigated the impact of varying the radius on the final segmentation performance, as shown in Fig \ref{fig:intlayers_PS}b. The segmentation performance remained consistent as the radius was increased between 20 and 50 voxels. However, when the radius was set to 10 voxels, a small percentage of the Jacobian determinants of the fine-tuning fields were non-positive, corresponding to 2/100 incorrect topologies. This is likely due to the decreased number of voxels in the prior representing the myocardium and the area enclosed by it, in combination with the additional smoothing modification that restrict the flexibility of the diffeomorphic fields.


 \section{Conclusion}
 We have shown that TEDS-Net is able to achieve highly accurate myocardium segmentations whilst ensuring topology preservation. We introduced additional diffeomorphic-encouraging modifications, which were found to play a crucial role in enforcing an one-to-one mappings in the generated discrete fields. Our method successfully segmented the myocardium in unaligned MRI heart slices, with different pathologies that had different thicknesses and circumferences, using the same general prior shape for all. This flexible, easy to train method has the potential to have a high impact in future clinical segmentation work.

\section*{Acknowledgments:}
MW and ND is supported by the Engineering and Physical Sciences Research Council (EPSRC) and Medical Research Council (MRC) [grant number EP/L016052/1]. MJ is supported by the National Institute for Health Research (NIHR) Oxford Biomedical Research Centre (BRC), and this research was funded by the Well- come Trust [215573/Z/19/Z]. The Wellcome Centre for Integrative Neuroimaging is supported by core funding from the Wellcome Trust [203139/Z/16/Z]. AN is grateful for support from the UK Royal Academy of Engineering under the Engineering for Development Research Fellowships scheme.

%
%

\bibliographystyle{splncs04}
\bibliography{bib}
\end{document}